\documentclass[pdflatex,sn-mathphys-num]{sn-jnl}%

\usepackage{graphicx}%
\usepackage{multirow}%
\usepackage{amsmath,amssymb,amsfonts}%
\usepackage{amsthm}%
\usepackage{mathrsfs}%
\usepackage[title]{appendix}%
\usepackage{textcomp}%
\usepackage{manyfoot}%
\usepackage{booktabs}%
\usepackage{algorithm}%
\usepackage{algorithmicx}%
\usepackage{algpseudocode}%
\usepackage{listings}%
\usepackage[table]{xcolor}

\begin{document}


\title{Structural determinants of soft memory in recurrent biological networks}


\author{\fnm{Maria Sol} \sur{Vidal-Saez}}

\author{\fnm{Jordi} \sur{Garcia-Ojalvo}}

\affil{\orgdiv{Department of Medicine and Life Sciences}, \orgname{Universitat Pompeu Fabra}, \orgaddress{\street{Barcelona Biomedical Research Park, Dr Aiguader 88}, \city{Barcelona}, \postcode{08003}, \country{Spain}}}

\abstract{Recurrent neural networks are frequently studied in terms of their information-processing capabilities.
The structural properties of these networks are seldom considered, beyond those emerging from the connectivity tuning necessary for network training. However, real biological networks have non-contingent architectures that have been shaped by evolution over eons, constrained partly by information-processing criteria, but more generally  by fitness maximization requirements.
Here we examine the topological properties of existing biological networks, focusing in particular on gene regulatory networks in bacteria.
We identify structural features, both local and global, that dictate the ability of recurrent networks to store information on the fly and process complex time-dependent inputs.}


\maketitle

\section{Biological computation via networks of dynamical elements}
\label{sec:intro}

One of the defining characteristics of living organisms is their ability to adapt to changing environments.
Cognition arising from the human brain is usually considered the culmination of this achievement.
Given that the information-processing capabilities of the brain are known to emerge from the networks of neurons that form it, it is not surprising that idealized representations of neural networks have been used for computational purposes for over 80 years \cite{mcculloch1943logical}.
The fact that abstract neuronal network models are able to perform complex information processing tasks is a reflection of the substrate independence that characterizes network computation \cite{barack2021two}.
This substrate independence not only has led to the current revolution in machine learning, but also is enabling us to extend the idea of network computation to other types of biological systems beyond the brain \cite{Gunawardena2022}.
A particular and relevant situation is that of individual (non-neural) cells, whose adaptability has been postulated to rely on complex networks of interactions among their biomolecular components, including (but not restricted to) genes and proteins at the level of transcriptional regulation \cite{lee_transcriptional_2002,martinez-antonio_identifying_2003}.
Interestingly, these gene regulatory networks (GRNs) are in some ways more accurate instantiations of artificial neural networks \cite{mjolsness1991connectionist} than the networks of biological neurons on which they were inspired. 

Mounting evidence in recent years has shown that individual cells perform complex computations. 
The bacterium \textit{Escherichia coli}, for instance, has been seen to exhibit temporal associative learning, relating pairs of stimuli that tend to occur together in a specific order, such as a rise in temperature followed by low oxygen conditions, which can be indicative of the microbe having been ingested by a mammal \cite{tagkopoulos_predictive_2008}.
A similar phenomenon has been observed in the fungal pathogen \textit{Candida albicans}, which can learn to respond to a glucose increase (indicative that the pathogen has entered a host) by upregulating genes responsible for oxidative stress resistance, as a way to preemptively react to the host's immune system \cite{schild_genes_2007}.

Learning the temporal structure of the environment, such as in the examples listed above, and processing time-dependent information in general, requires memory.
In that context, studies have shown that bacteria display both short- and long-term memory. Specifically, carefully designed culture experiments have revealed that the stress response of \textit{Bacillus subtilis} cells relies not only on their present growth conditions, but also on their environmental history \cite{wolf_memory_2008}.
While classical approaches to memory are usually based on hard-wiring network connections using Hebbian rules \cite{brown_legacy_2003}, recent work as shown that memory can also be \textit{soft-wired} through generalized chaos synchronization \cite{Casal2020}.
To that end, the network must exhibit self-sustained dynamical behavior, which naturally results from recurrent connections \cite{sussillo2009generating}.
Such recurrent networks are indeed known to process time-dependent information, in what is known as state-dependent computations \cite{buonomano_state-dependent_2009}.
A particular type of computational paradigm in that respect is based on the concept of reservoir computing \cite{verstraeten_experimental_2007}, which was initially proposed almost simultaneously in the fields of machine learning and computational neuroscience, implemented by computational systems that came to be known as echo state networks \cite{jaeger_echo_2001} and liquid state machines \cite{maass_real-time_2002}, respectively.

Following previous work \cite{jones_is_2007,gabalda-sagarra_recurrence-based_2018}, here we propose that the inherent structure and dynamics of gene regulatory networks allow bacteria to process temporal information.
Our aim is to establish how the structural properties of the network, both global and local, determine its computational capabilities, focusing in particular on its ability to store information from the environment on the fly (what we term \textit{soft memory}).
Specifically, in what follows we first study the effect of the global network dynamics on its memory capacity. 
We then examine if a balance between activation and repression is needed to have a functional network (similarly to what happens in neuronal networks \cite{van1996chaos}). 
Next we analyze whether the local topological properties of biological gene regulatory networks contribute to their memory encoding. Finally, we search for smaller set of genes within the \textit{E. coli}'s network that can perform well in memory-demanding tasks, which could be helpful for the experimental validation of this computational paradigm in single cells.

\section{Global structural determinants of soft memory}

\begin{figure}[htbp]
    \centering
    (a)
\begin{minipage}[t]{0.4\textwidth}
\vspace{0pt}
\includegraphics[width=\textwidth]{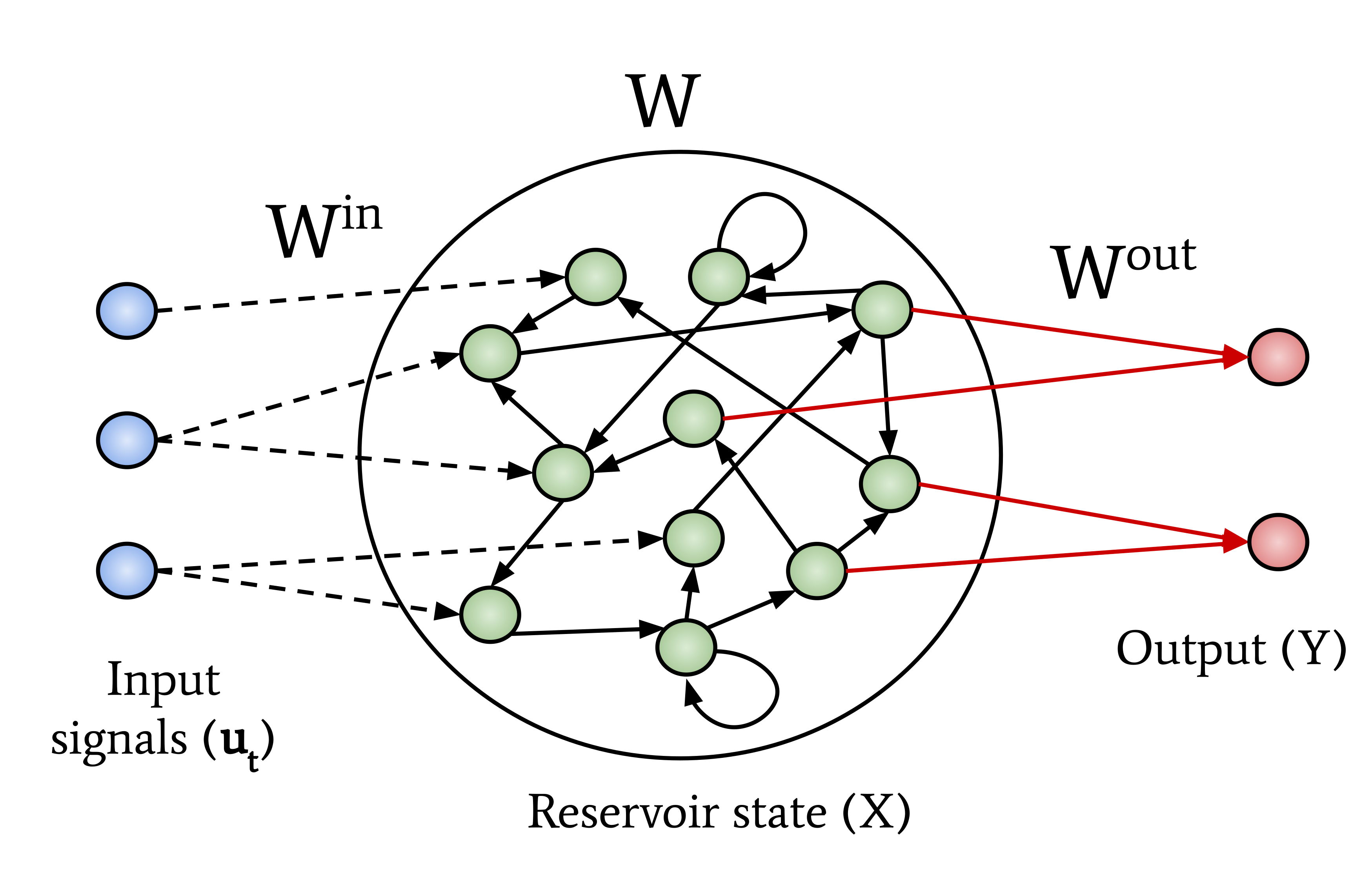}
\end{minipage}
(b)
\begin{minipage}[t]{0.5\textwidth}
\vspace{0pt}
\includegraphics[width=\textwidth]{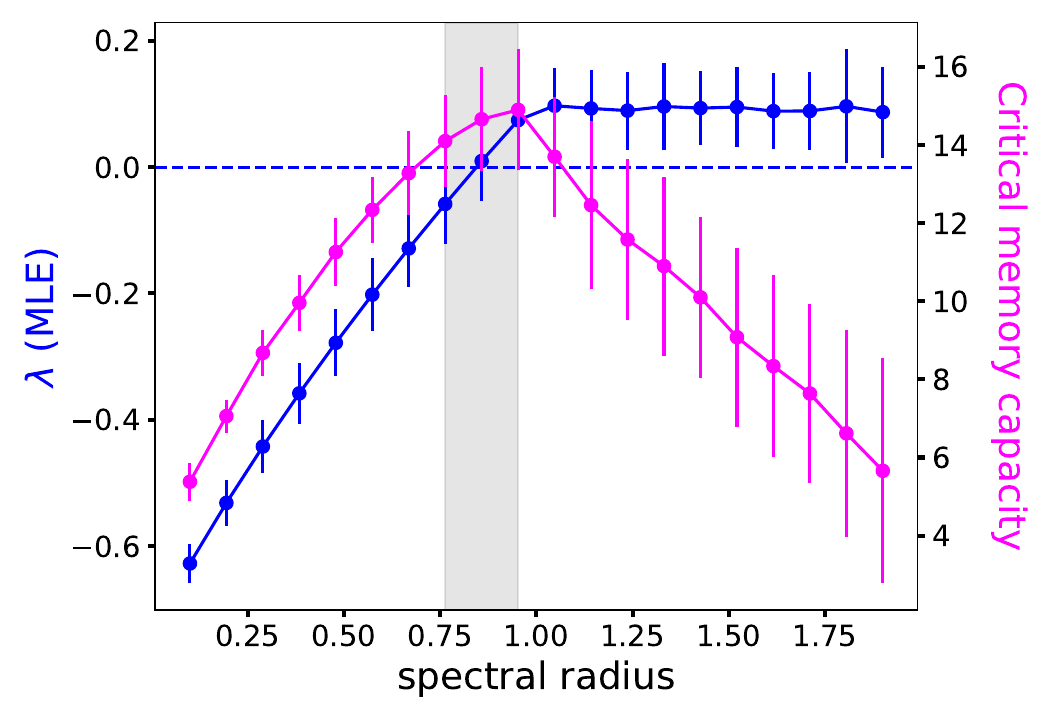}
\end{minipage}
(c)
\begin{minipage}[t]{0.45\textwidth}
\vspace{0pt}
\includegraphics[width=\textwidth]{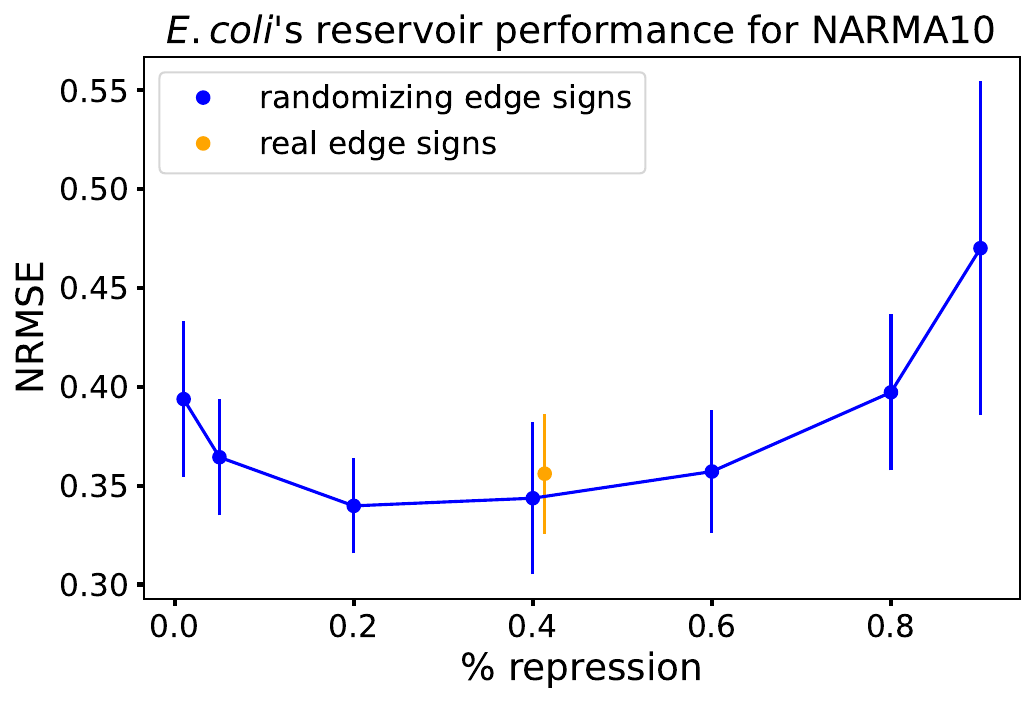}
\end{minipage}
(d)
\begin{minipage}[t]{0.45\textwidth}
\vspace{0pt}
\includegraphics[width=\textwidth]{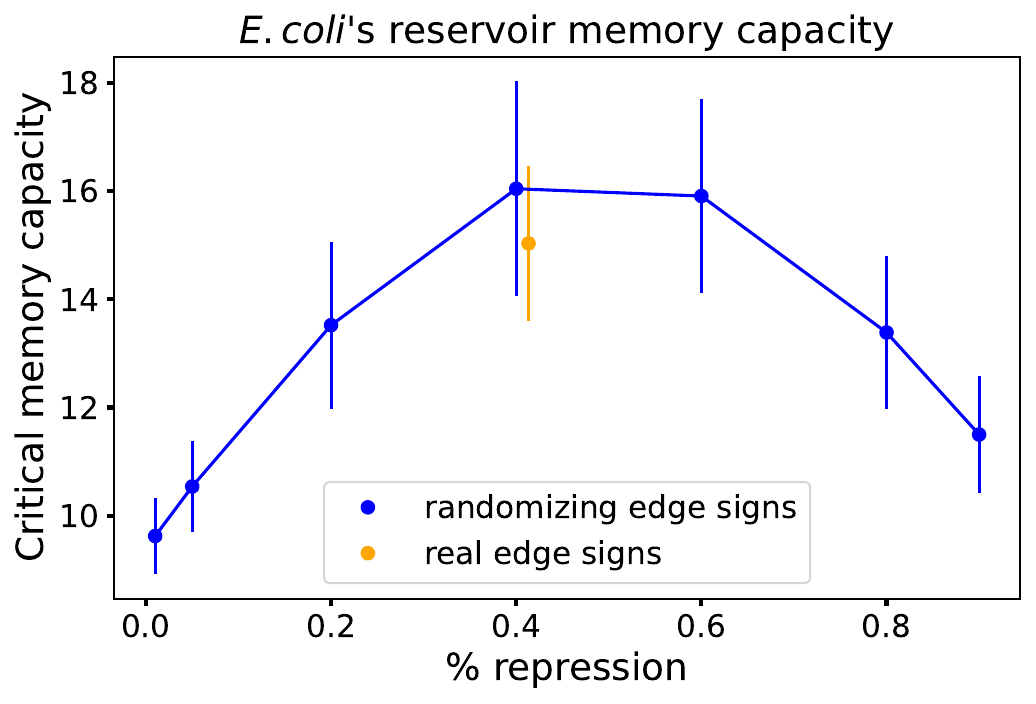}
\end{minipage}
\caption{(a) Schematic representation of a reservoir computing architecture. The input nodes in blue denote
the external signals acting upon the reservoir nodes. The recurrent core is represented by the green nodes.
The output nodes (red) form a feedforward subgraph that reads the information encoded in the reservoir to make decisions. Only the readout weights $W^{out}$ (red arrows) are trained so that the output $Y$ approximates target output signal(s).
(b) Maximum Lyapunov exponent (blue) and critical memory capacity (magenta) as a function of the spectral radius of the reservoir. The maximal critical memory capacity occurs in the vicinity of the transition from ordered to chaotic dynamics (when $\lambda\approx 0$, horizontal dashed line in blue).
The values shown are the mean over 50 realizations of the network with different random input and recurrent weights. Errors were estimated as the standard deviation over those 50 realizations.
(c) Network performance for a 10th-order NARMA task and (d) critical memory capacity as a function of the percentage of repressing nodes in randomized networks, while the real network results are shown in orange. The values shown are the mean over 100 realizations. Errors were estimated as the standard deviation over those 100 realizations. }
\label{fig:global}  
\end{figure}

The gene regulatory network of \textit{E. coli} has been extensively documented and made publicly available in resources such as EcoCyc \cite{keseler_ecocyc_2011}.
Using the EcoCyc database, we previously established that the architecture of this biological network fits within the paradigm of reservoir computing \cite{gabalda-sagarra_recurrence-based_2018}.
Specifically, the network has a core sub-structure formed by a relatively small number of genes that are recurrently connected to each other, known as the reservoir (Fig.~\ref{fig:global}a).
This recurrent core receives input signals, shown in blue in the figure, and projects their dynamics nonlinearly into a high-dimensional space, which enables further on-the-fly classification of those dynamical signals by a downstream readout layer (red nodes in the figure). 
As shown in Table~\ref{tab:properties}, the recurrent core is a small fraction of the entire network, yet it endows the system with notable computational properties, while simultaneously admitting training procedures that are biologically feasible \cite{gabalda-sagarra_recurrence-based_2018}, as we discuss below.

\begin{table}[htbp]
   \centering
   \begin{tabular}{c|c|c|c}
      \rowcolor{gray!45}   & Nodes & Edges & Mean degree \\
      \hline
       \cellcolor{gray!45!}  Whole graph & 3236 & 8366 & 5.17 \\
       \cellcolor{gray!45!}  Recurrent core & \cellcolor{gray!15!} 70 & \cellcolor{gray!15!} 317 & \cellcolor{gray!15!} 9.05 
   \end{tabular}
   
   \vspace{2mm}
   \caption{Structural properties of \textit{E. coli}'s gene regulatory network.}
   \label{tab:properties}
\end{table}

\subsection{Network simulations}
sol
The dynamics of the recurrent core can be simulated using a temporally discrete update rule given by:
\begin{equation} 
    \mathbf{x}_{t}  = \tanh(W^{in}\mathbf{u}_t + W\mathbf{x}_{t-1}),
\label{res_dyn}
\end{equation}
where $t=1,\ldots,T$ are the discrete time points in the training dataset, $\mathbf{x}_{t}\in \mathbb{R}^{N\times 1}$ is the state vector of the reservoir units (with $N$ representing the number of reservoir nodes) and $\mathbf{u}_t\in \mathbb{R}^{N_{in}\times 1}$ is the input signal vector (with $N_{in}$ denoting the number of inputs), both at time $t$.
The weight matrices $W^{in}\in \mathbb{R}^{N\times N_{in}}$ and $W\in \mathbb{R}^{N\times N}$ represent the input-reservoir connections and the recurrent connections within the reservoir, respectively.
The elements of $W^{in}$ are randomly chosen to be either -0.05 or 0.05.
At the same time, the elements of $W$ are real random numbers drawn from a uniform distribution $\mathcal{U}$(0,1) if the link exists (i.e. if a protein and its corresponding target gene are connected in the EcoCyc database), and 0 otherwise.
The sign of the non-zero elements of $W$ is set according to whether the regulation is an activation or a repression.
Additionally, the $W$ matrix is normalized to have a given spectral radius.
A spectral radius close to, but below, 1 gives rise to the so-called \textit{echo-state property} \cite{lukosevicius_reservoir_2009,jaeger_echo_2001}, which we will discuss in some detail in what follows. 

The dynamics of the reservoir is then fed into the readout nodes, which define the output as a weighted sum of the state of the nodes from which it receives information:
\begin{equation}
    Y  =W^{out}X
    \label{output}
\end{equation}
Here $Y \in \mathbb{R}^{N_{out}\times T}$ is a matrix with all predicted outputs over time (with $N_{out}$ denoting the number of readout nodes/outputs), $X \in \mathbb{R}^{N\times T}$ is a matrix that contains the states of the reservoir nodes over time, and $W^{out} \in \mathbb{R}^{N_{out}\times N}$ is the weight matrix for the reservoir-readout connections.
In the reservoir computing paradigm, training the network corresponds to fitting the $W^{out}$ weights so that the network output approximates the target outputs of the training dataset.
A simple training method used in machine learning is ridge regression, where the $W^{out}$ weights are obtained from: 
\begin{equation}
    W^{out} = Y^{\mathrm{target}}X^{\mathsf{tr} }(XX^{\mathsf{tr} } + \gamma ^{2}I)^{-1}
\label{ridge}
\end{equation}
Here $Y^{\mathrm{target}}\in \mathbb{R}^{N_{out}\times T}$ is a matrix with the
target outputs over time and $X^{\mathsf{tr} }$ is the transpose of $X$. $I$ is the identity matrix, and the regularization coefficient $\gamma$ is introduced to avoid overfitting \cite{lukosevicius_reservoir_2009}. 
Ridge regression is a variation of the least squares method that penalizes regression coefficients with large absolute values.
In doing so it introduces a certain bias, but on the other hand it also reduces the variance of the estimate.
This allows estimating the linear regression parameters when the predictor variables are
strongly correlated, making it a common readout choice in the context of reservoir computing \cite{wyffels_stable_2008}. 
On the other hand, it is not a realistic training method for biological networks.
In that case, genetic algorithms mimicking the learning role of evolution can be used \cite{gabalda-sagarra_recurrence-based_2018,watson2016can}.

To be functional, recurrent cores need to exhibit particular design properties.
In particular, to combine both separability and generalizability, they need to operate between ordered and chaotic dynamics, i.e. near the so-called \textit{edge of chaos} \cite{bertschinger_real-time_2004,legenstein_edge_2007,Morales2021,vidal_BBRC_24}.
This refers to a critical state between ordered
dynamics (where perturbations quickly die out into an attractor) and chaotic dynamics (where perturbations are amplified).
This property is particularly interesting, because of existing evidence indicating that cortical circuits are close to criticality \cite{beggs_criticality_2007,beggs_neuronal_2003,r_chialvo_critical_2004}.
There is so far no direct evidence suggesting that gene regulatory networks operate near criticality (but see \cite{criticality_blai} for a synthetic implementation of critical behavior in \textit{E. coli}).
Since we are interested in memory encoding in these types of networks, we measure in what follows the memory capacity of \textit{E. coli}'s reservoir as it undergoes a phase transition from ordered to chaotic dynamics.
With this analysis we seek to establish if the dynamical regime of the reservoir has an effect on its memory capacity.

To control the phase transition from ordered to chaotic dynamics we tune the spectral radius $\rho$ of the adjacency matrix representing the recurrent core of \textit{E. coli}'s gene regulatory network.
This quantity corresponds to the largest absolute value of all eigenvalues of the matrix, and determines the dynamical stability of the reservoir when no input is fed into the network \cite{lukosevicius_reservoir_2009}.
Large spectral radius values can lead to multiple fixed points, or to periodic or even chaotic
(for large enough nonlinearity) attractor modes.
On the other hand, a spectral radius around 1 leads to the so-called \textit{echo state property}, which ensures that the effect of initial conditions on the reservoir state fades away asymptotically in time \cite{jaeger_short_2001,jaeger_echo_2001,  yildiz_re-visiting_2012}.
Taking this into account, we have tuned the spectral radius value in order to obtain reservoirs with ordered as well as chaotic dynamics. 

\subsection{Critical memory capacity}\label{subsec:memcap}

To analyze the memory capacity of \textit{E. coli}'s reservoir we have chosen a variation of a common benchmark task, namely the memory capacity \cite{jaeger_short_2001}, as a quantifier of the degree of memory exhibited by the network. 
In this task, a single input node feeds  into the reservoir a signal $u_{t}$ drawn from a random uniform distribution $\mathcal{U}$(-1,1). A readout node is then trained to produce an output that matches a delayed version of the input signal. 
To evaluate the short-term memory capacity of the network, we compute the normalized covariance between the delayed input and the output:
\begin{equation}
    MC_k = \frac{\mathrm{cov}^{2}(u_{t-k}, y_t)}{\sigma ^{2} (u_{t-k})\cdot \sigma ^{2}(y_t)}. 
    \label{MCk}
\end{equation}
We define the critical memory capacity $k^*$ of the network as the maximum delay $k$ that fulfills $MC_k>0.5$. Similar approximations to the short-term memory capacity have been made in other works (see e.g. \cite{boedecker_information_2012}).
In order to measure $MC_k$ for each $k$, we generated 10 input-output series (input: $u_{t}$, output: $u_{t-k}$) of 1000 steps each, using 9 of them for the training phase of the ridge regression, and the remaining one to test its performance by measuring the corresponding $MC_k$, as given by Eq.~\eqref{MCk}.

\subsection{Measuring chaos in reservoirs}
To keep track of the transition from order to chaos we have used the maximum Lyapunov exponent (MLE), a hallmark indicator of criticality. This indicator
quantifies the exponential divergence of two initially close trajectories of a dynamical system in state space \cite{derrida_random_1986}, and has been already applied in the reservoir computing framework \cite{legenstein_edge_2007, bertschinger_real-time_2004, boedecker_information_2012}.
It is defined by:
\begin{equation}
    \lambda = \lim_{k \to \infty }\frac{1}{k} \mathrm{ln}\left(\frac{\gamma _k}{\gamma _0}\right),
\end{equation}
where $\gamma _0$ is the initial distance between the two trajectories, and $\gamma _k$ is the distance at time $k$.
Chaotic dynamics is typically associated with a positive MLE ($\lambda>0$), while for sub-critical systems (ordered phase) $\lambda< 0$.
A transition thus occurs at $\lambda \approx 0$.
We estimate the MLE following the method described in Ref.~\cite{Sprott2003-ln}. 

\subsection{Maximal memory capacity at the edge of chaos}
\label{sec:mem_cap}
Using the approaches described in the previous sections, we now compute the critical memory capacity of the \textit{E. coli}'s recurrent core and the MLE of its dynamics, for increasing values of the spectral radius $\rho$.
These results are displayed in Fig.~\ref{fig:global}(b). The figure shows that the system transitions from a sub-critical ($\lambda <0$) to a super-critical ($\lambda >0$) state at a spectral radius $\rho \approx 0.9$. 
In turn, the maximal critical memory capacity of the network is reached just above the onset of chaos, when $\rho \approx 0.95$. Thus, the GRN of \textit{E. coli} has maximal memory at the vicinity of the order-to-chaos transition.

The spectral radius $\rho$ of the GRN could be tuned biologically by altering globally the gene expression potential of cells (varying in a controlled manner the levels of resource cellular components such as RNA polymerases, housekeeping sigma factors, or ribosomes).
The model thus makes an experimental prediction: the memory capacity of \textit{E. coli}'s GRN should depend non-monotonically on its global expression strength, according to the results of Fig.~\ref{fig:global}(b). 

\subsection{Activation/repression balance in GRNs}

The Ecocyc dataset \cite{keseler_ecocyc_2011} includes information about the sign of all transcriptional interactions, namely whether they are activations or repressions.
The proportion of repressing interactions in the entire \textit{E. coli} GRN is 24\%, while in the recurrent core is 41\%.
We then ask whether this activation/repression ratio has an effect on the performance of the network.
In other words, has evolution tuned the activation/repression ratio in natural reservoirs to operate near optimal performance? Is the memory capacity of the network affected if we modify this ratio?
This idea has been explored in the context of neuronal networks \cite{Casal2020}, but not in GRNs.
In the neuronal case, the excitatory/inhibitory (E/I) balance, defined as the balance between excitation and inhibition of synaptic activity in a neuronal network, plays a crucial role in maintaining the regular functionality of the brain \cite{deco_how_2014}.
This equilibrium, responsible for regulating normal spike rates, is disrupted in numerous pathological conditions, resulting in either excessive or diminished excitation relative to inhibition, termed E/I imbalance \cite{ghatak_novel_2021,kirischuk_keeping_2022}.

To study the effect of the activation/repression balance in gene regulatory networks, we randomized the signs of the reservoir edges, while maintaining a proportion of inhibition.
The performance of the \textit{E. coli}'s reservoir was studied for increasing repression/activation ratios, using the 10th-order nonlinear autoregressive moving average (NARMA) task, a memory-demanding benchmark commonly used in the context of neural networks \cite{jaeger_adaptive_2002}. The task consists in training a network to reproduce the output of the 10th-order NARMA system \cite{atiya_new_2000}, a discrete dynamical system with input values $s_t$ drawn from a random uniform distribution $\mathcal{U}$(0, 0.5), while the output $y^{\mathrm{NRM}}(t)$ is defined by:
\begin{equation}
    y^{\mathrm{NRM}}(t+1) = 0.3y^{\mathrm{NRM}}(t) + 0.05y^{\mathrm{NRM}}(t)\sum_{i=0}^{9}y(t-i)+1.5s(t-9)s(t) + 0.1
    \label{narmaeq}
\end{equation}

To test if the dynamics of \textit{E. coli}'s reservoir can represent the temporally correlated NARMA input, the network was simulated with a single input node feeding the $s(t)$ series into the system. 
A readout node was then trained to reproduce the output $y^{\mathrm{NRM}}(t)$ of the 10th-order NARMA system using only the instantaneous state of the network.
For each realization, 10 NARMA series of 1000 steps were generated, using 9 of them for the training phase and the remaining one to test the performance. 
The main challenge of the 10th-order NARMA task is that the output of the time series depends on the input and output values of the last 10 time steps. This information about the past must be encoded in the reservoir state for the predicted output to be able to accurately model the input. 

To quantify the network performance, we used the normalized root mean squared Error (NRMSE) between the
predicted and target output signals, defined as
\begin{equation}
    NRMSE = \sqrt{\frac{\left \langle (y(t) - y^{\mathrm{target}}(t))^{2} \right \rangle_{t}}{\left \langle (y^{\mathrm{target}}(t) - \left \langle y^{\mathrm{target}}(t) \right \rangle_{t})^{2} \right \rangle_{t}}}, 
\label{NRMSE}
\end{equation}
where $y(t)$ is the output predicted by the readout, $y^{\mathrm{target}}(t)$ is the target output ($y^{\mathrm{NRM}}(t)$ in this case), and $\left \langle \cdot \right \rangle_{t}$ indicates the average over time.

We now challenge the \textit{E. coli}'s reservoir with the NARMA task described above, for varying proportions of activation/repression. The results obtained are displayed in Fig.~\ref{fig:global}(c).
The results shows that the performance of the \textit{E. coli}'s reservoir for the NARMA task varies non-monotonically with respect to the repression ratio.
{This behavior also holds for the critical memory capacity task, as defined in Sec.~\ref{subsec:memcap} and shown in Fig. \ref{fig:global}d.
In particular, the best performance is observed when the percentage of repression is within the range 40\%-60\%, which encompasses the value exhibited by the biological dataset \cite{keseler_ecocyc_2011}, shown by the orange dot in Figs.~\ref{fig:global}(c,d).} In other words, the biological reservoir operates near the activation/repression ratio that yields optimal performance. 
In summary, an optimal balance between activation and repression is needed in the \textit{E. coli}'s reservoir in order to endow it with temporal information processing capabilities. A similar result has been reported in the nervous system of the nematode \textit{C. elegans} \cite{Casal2020}.

\section{Effect of local topology on memory capacity}

In the previous section we analyzed the impact of the dynamical regime of the \textit{E. coli}'s reservoir on its memory capacity, following previous studies on artificial recurrent networks that showed improved performance at the edge of chaos \cite{legenstein_edge_2007}.
We now take a different perspective: does the \emph{local} topology of biological reservoirs have an influence on their memory capacity? 

\subsection{Memory motifs}

The connectivity patterns of \textit{E. coli}'s GRN are not random; they have been actively shaped by natural selection \cite{alon_introduction_2019}.
To look for \textit{meaningful} patterns, we can compare the real network to an ensemble of randomized networks, with the same number of nodes and edges as the real network, but where the
connections between nodes are made at random.
Patterns that occur in the real network significantly more often than in randomized networks are called \emph{network motifs} \cite{milo_network_2002,shen-orr_network_2002}.
We focus in what follows on three circuits: self-loops, mutual regulation circuits and feedforward loops (FFLs), where the interactions can correspond to either activation or repression. Self-loops and FFLs are network motifs of \textit{E. coli}'s GRN \cite{alon_introduction_2019}. Mutual regulation circuits have been included, since we hypothesize that some local recurrence is needed to have a network with memory. From now on we will refer to the these three circuits as \textit{memory motifs}.
{We note that these are the smallest circuits that contain feedback and feedforward interactions. While only one node is needed to have a feedback loop, three nodes are needed for a feedforward connection. The mutual regulation architecture, in turn, is the second simplest topology that exhibits feedback. We choose this type of circuit to go beyond the somewhat singular character of single-gene circuits.}

Table \ref{tab:reservoir} shows the number of copies of each memory motif that are embedded in the whole gene regulatory network and in the recurrent core of \textit{E. coli}, compared with the expected number of motifs in their correspondent randomized networks.
As shown in the table, the normalized relative abundance of FFL circuits relative to the randomized networks, as measured by the z-score, is reduced by a factor of over 150 between the whole graph and the reservoir, indicating that the whole network is much more enriched with FFLs than its recurrent core.
This result makes sense considering that the whole graph comprises both the reservoir and the readout. The latter is a feedforward architecture with 98\% of the nodes, which is locally enriched with feedforward loops (FFL).
The z-scores of the self-loop and mutual regulation circuits, on the other hand, are reduced only by a factor of around 3 between the whole graph and the reservoir, although both circuits are significantly relevant motifs (i.e. the z-score is significantly larger than 1) in both networks.
Interestingly (although expectedly), all 28 mutual regulation motifs of the whole graph are part of the reservoir.

\begin{table}[htbp]
    \centering
    \begin{tabular}{ c|c|c|c}
        \rowcolor{gray!45} \textbf{Whole graph} & Real network & Randomized  network & z-score \\
         \cellcolor{gray!45!} self-loop & 126 & 3 $\pm$ 2 & 79 \\
         \cellcolor{gray!45!} mutual regulation & \cellcolor{gray!15!} 28 & \cellcolor{gray!15!} 3 $\pm$ 2 & \cellcolor{gray!15!} 14\\
         \cellcolor{gray!45!} FFL & 4798 & 17 $\pm$ 4 & 1167
    \end{tabular}
   
   \vspace{2mm}
        \begin{tabular}{ c|c|c|c}
        \rowcolor{gray!45} \textbf{Reservoir} & Real network & Randomized  network & z-score \\
         \cellcolor{gray!45!} self-loop & 55 & 5 $\pm$ 2 & 26 \\
         \cellcolor{gray!45!} mutual regulation & \cellcolor{gray!15!} 28 & \cellcolor{gray!15!} 10 $\pm$ 3 & \cellcolor{gray!15!} 5.8\\
         \cellcolor{gray!45!} FFL & 147 & 73 $\pm$ 10 & 7.6
         \end{tabular}
   \vspace{2mm}

    \caption{Number of motifs in the entire GRN (top) and recurrent core (bottom) of \textit{E. coli} compared to the expected number in randomized networks with the same number of nodes and edges (mean $\pm$ s.d. computed over 1000 realizations). The z-score is computed as z-score$=  \frac{N_{\mathrm{real}} - \left \langle N \right \rangle_{\mathrm{rand}}}{\sigma_{\mathrm{rand}} }$.}
    \label{tab:reservoir}
\end{table}

\subsection{Memory capacity of the memory motifs}\label{subsec:mem_cap_sub}

To assess the relation of the three memory motifs introduced above with the memory capacity of the recurrent core, we  sampled $\approx 2400$ sub-reservoirs of different sizes from the whole \textit{E. coli}'s reservoir of 70 genes. To do it, we started by removing nodes from the reservoir sequentially and at random. We then pruned the resulting network to make sure it is still a reservoir \cite{gabalda-sagarra_recurrence-based_2018}, and kept only its giant connected component.
We iterated this process until we reached a reservoir of only 3 nodes (minimum size considered).
We repeated this process 100 times starting from different nodes, and ended up with a total of 2394 sub-reservoirs. 
This procedure led to reservoirs with sizes from 3 to 69 nodes. 

\begin{figure}[htbp]
\centering
(a)
\begin{minipage}[t]{0.23\textwidth}
\vspace{0pt}
\includegraphics[height=\textwidth]{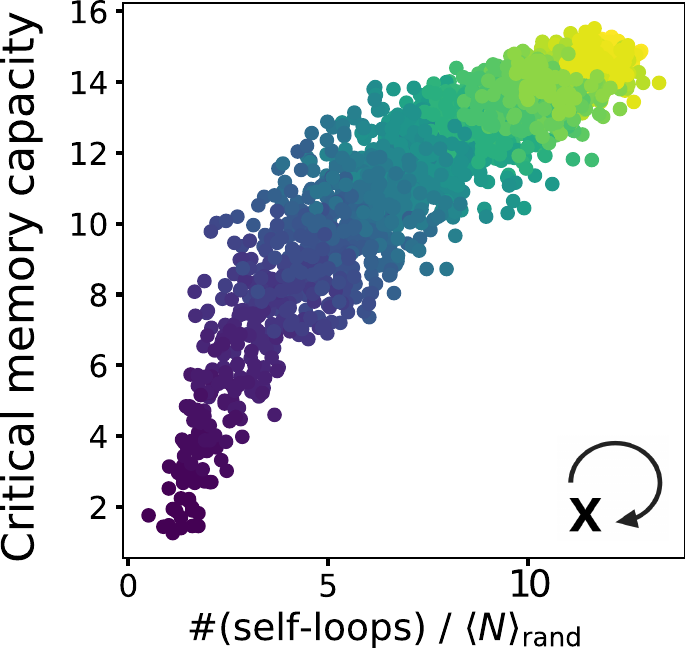}
\end{minipage}
~~(b)
\begin{minipage}[t]{0.23\textwidth}
\vspace{0pt}
\includegraphics[height=\textwidth]{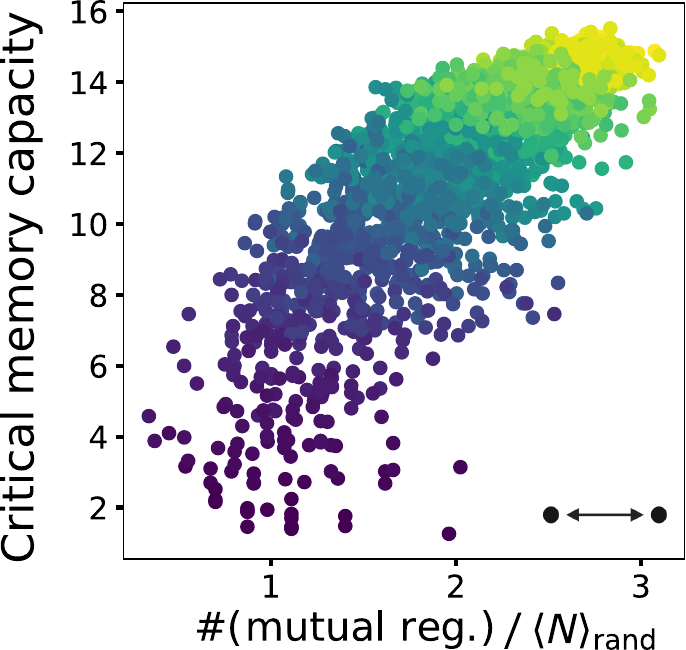}
\end{minipage}
\vspace{5pt}
~~~(c)
\begin{minipage}[t]{0.23\textwidth}
\vspace{1pt}
\includegraphics[height=\textwidth]{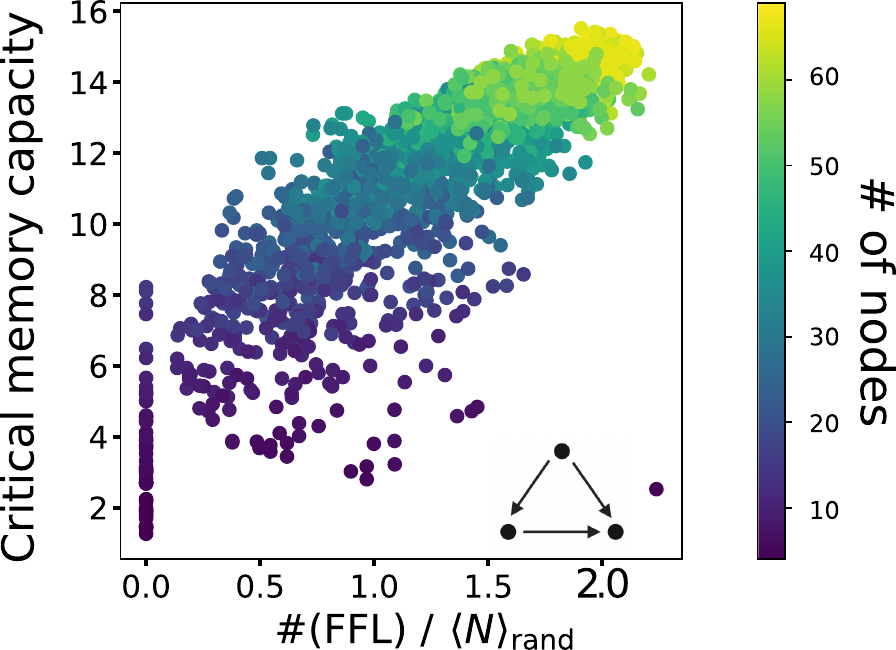}
\end{minipage}
\\
(d)
\begin{minipage}[t]{0.27\textwidth}
\vspace{0pt}
\includegraphics[width=\textwidth]{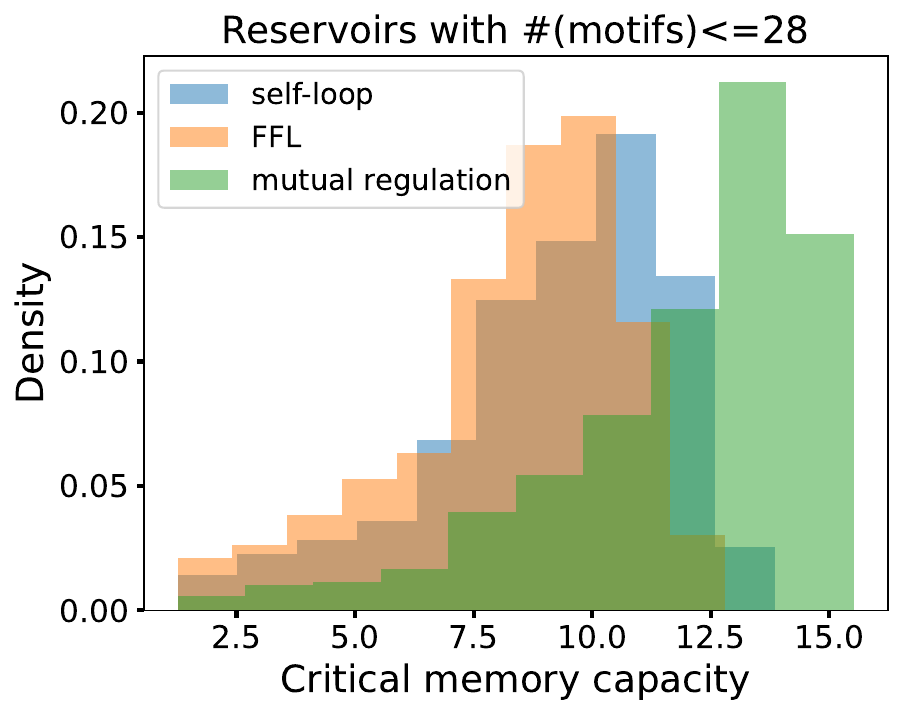}
\end{minipage}
(e)
\begin{minipage}[t]{0.27\textwidth}
\vspace{0pt}
\includegraphics[width=\textwidth]{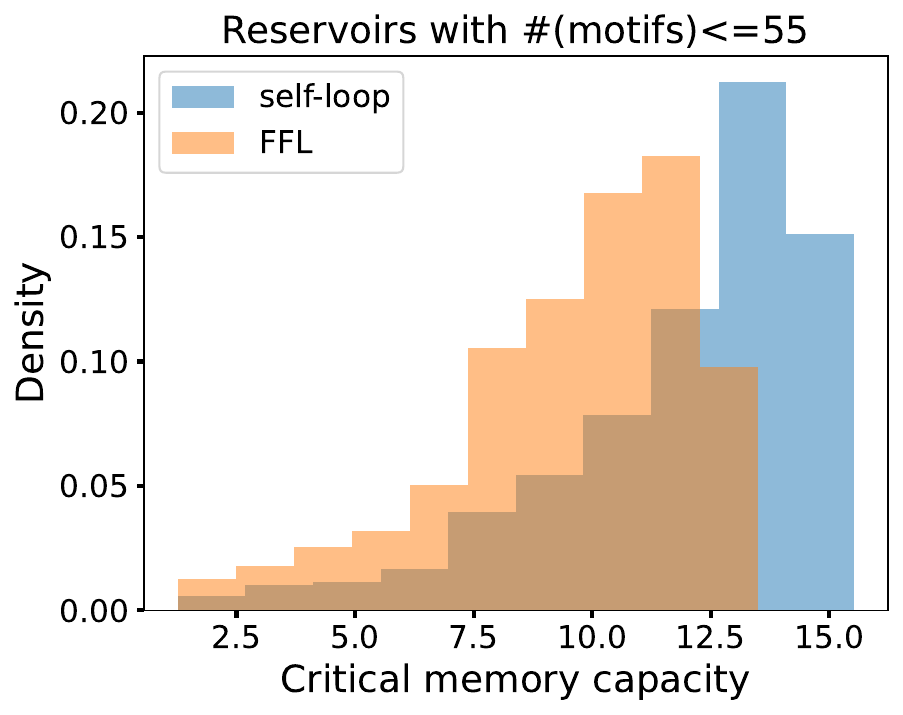}
\end{minipage}
\caption{
(a-c) Critical memory capacity vs number of memory motifs normalized by expected number in randomized networks.
The x axis in each case corresponds to the number of self-loops (a), number of mutual regulation circuits (b), and number of FFLs (c).
The critical memory capacity for each reservoir was computed as the average over 50 realizations.
The color coding represents the number of nodes in each reservoir (from 3 to 69).
The expected number of motifs in the random networks was calculated for networks with the same number of nodes and edges (mean over 1000 realizations).
(d,e) Critical memory capacity distributions for reservoirs with less motifs than an given threshold, for two threshold values.
The critical memory capacity for each sub-reservoir was computed as the average over 50 realizations.}
\label{fig:memory}  
\end{figure}

We then measured the critical memory capacity of the sub-reservoirs generated as described above.
The results are shown in Fig.~\ref{fig:memory}(a-c), as a function of the number of memory motifs, normalized by their expected number in randomized networks.
In these figures, each symbol represents a sub-reservoir, with its color denoting the number of nodes of each circuit.
To normalize the motif number, for each sub-reservoir we simulated its correspondent random network with the same quantity of nodes and edges. We counted the number of self-loops, mutual regulation circuits and FFLs in the random network. We repeated the process 1000 times, and then computed the expected number as the mean of the 1000 realizations.

The results shown in panels in Fig.~\ref{fig:memory}(a-c) reveal an overall increase of the memory capacity with the number of motifs for the three circuits. 
However, the quantitative dependence on the number of motifs is markedly different in the three cases.
Specifically, the sub-reservoirs are more enriched with self-loops than mutual regulation and FFL motifs, as reflected in the fact that the x-axis range is much larger in panel (a) than in (b,c).
This is consistent with the results presented in Table \ref{tab:reservoir}. Another interesting feature of Fig.~\ref{fig:memory}(c) is that FFLs do not determine the reservoir's memory, since even networks with no FFLs whatsoever have reservoirs with memory capacity in a broad range (from 0 to 8). Clearly, these networks have other motifs that give them memory. It is worth noting that this broad range of memory capacity levels in the absolute absence of FFL motifs does not exist for self-loop or mutual regulation motifs.

A comparison between panels (a) and (b) of Fig.~\ref{fig:memory} also reveals that mutual regulation circuits are much more efficient than self-loops in enhancing memory.
Specifically, panel (b) shows that maximum memory capacity ($\sim 15$) can be reached with a normalized number of mutual regulation circuits of around 3, whereas these number grows to around 10 for self-loops (panel a).
This is also reflected in the memory capacity distributions for the three memory motifs shown in Fig.~\ref{fig:memory}(d), which is restricted to sub-reservoirs with less than 28 motifs.
When this threshold is increased to 55 (panel e of the figure), the difference between self-loops and FFLs also becomes evident.

\section[Smaller \textit{memorious} GRNs] {Smaller \textit{memorious} GRNs }

We now aim to identify the smallest sub-networks of the \textit{E. coli}'s reservoir that perform well in memory-demanding tasks.
To that end, we confronted the 2394 sub-reservoirs introduced in the previous section with both the memory capacity and the 10th-order NARMA tasks defined above, and also with a biologically inspired task, namely a delayed AND integration.
We then looked for networks with a good size-performance trade-off across the three tasks. 
We first describe the delayed AND task. 

\subsection{Delayed AND task}

\begin{figure}[htbp]
\centering
(a)
\begin{minipage}[t]{0.7\textwidth}
\vspace{0pt}
\includegraphics[width=\textwidth]{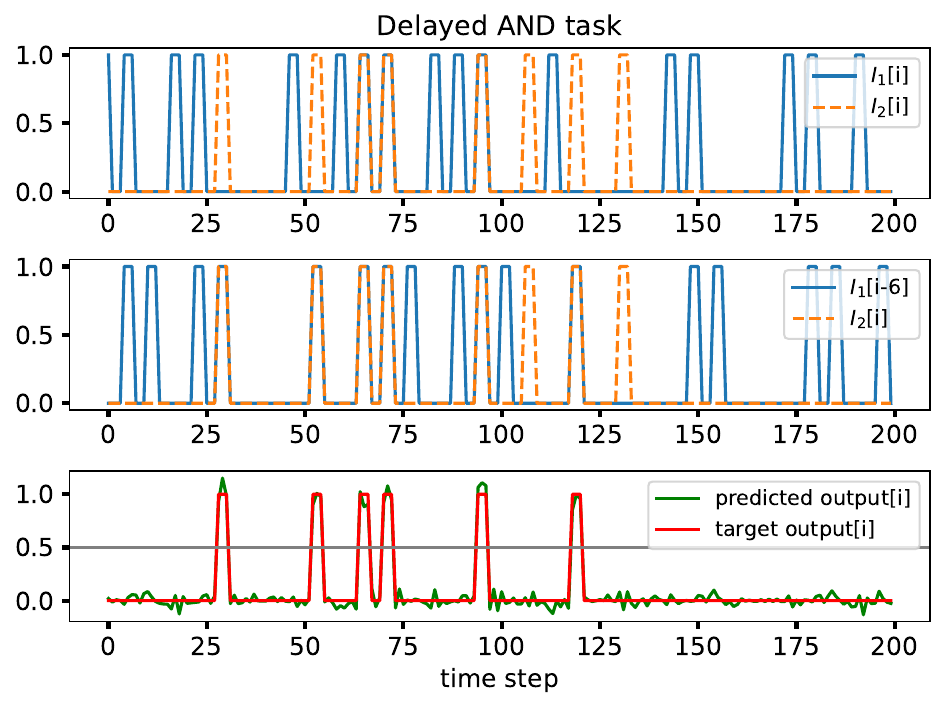}
\end{minipage}
\\
\begin{minipage}[t]{0.3\textwidth}
(b)
\vspace{0pt}
\includegraphics[width=\textwidth]{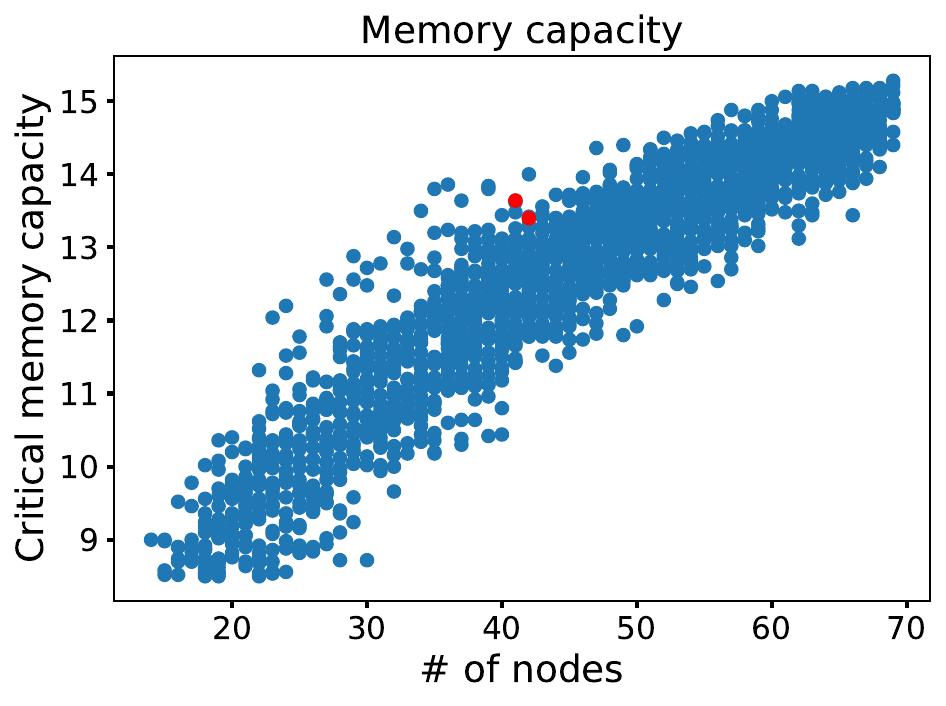}
\end{minipage}
\begin{minipage}[t]{0.3\textwidth}
(c)
\vspace{-2pt}
\includegraphics[width=\textwidth]{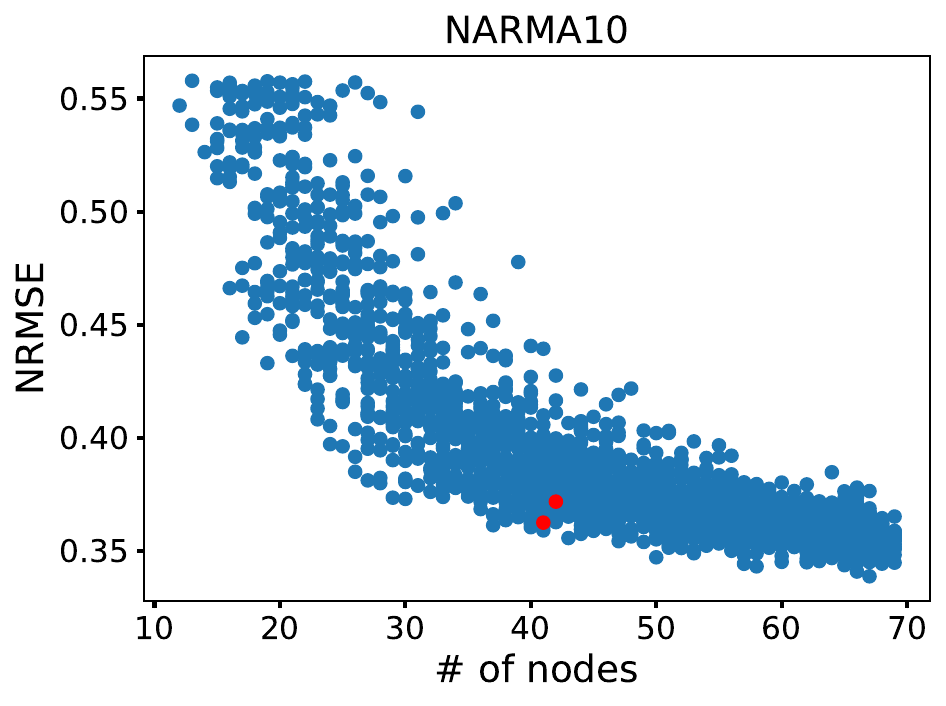}
\end{minipage}
\begin{minipage}[t]{0.3\textwidth}
(d)
\vspace{-2pt}
\includegraphics[width=\textwidth]{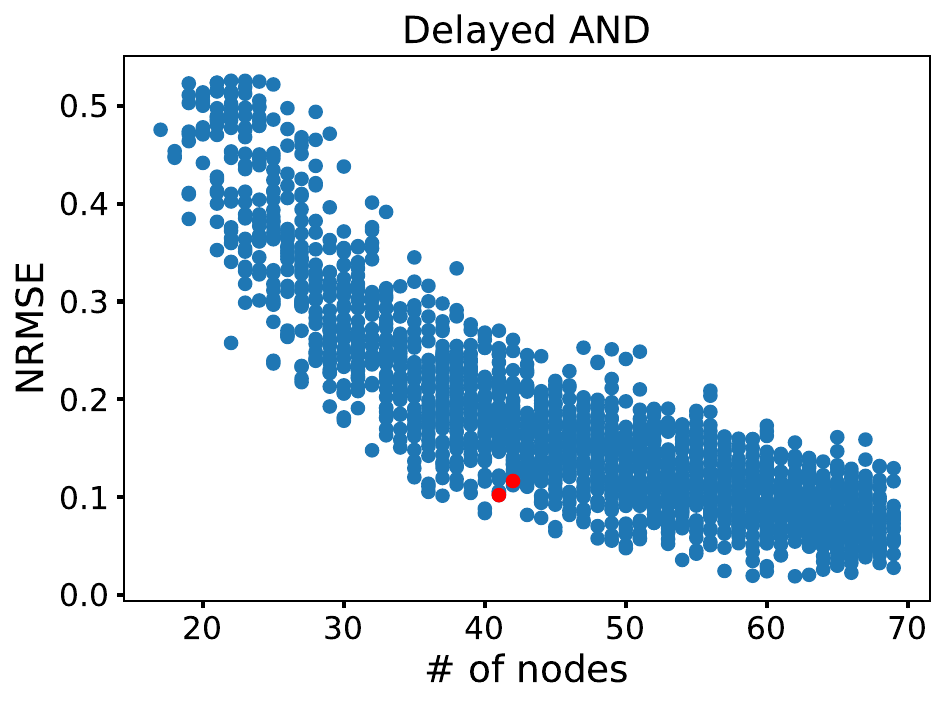}
\end{minipage}
\caption[A delayed AND task]{
(a) A delayed AND task.
Top row: time series of the two input signals ($I_1$[i] and $I_2$[i]). Middle row: the input $I_1$ delayed 6 time steps ($I_1$[i-6]) in relation to input $I_2$[i]. Bottom row: the target output (red) at time step $i$ is calculated as an AND gate between $I_1$[i-6] and $I_2$[i]. In green the predicted output by the whole \textit{E. coli}'s reservoir is shown.
(b-d) \textit{E. coli}'s sub-reservoirs performance in terms of network size, for the critical memory capacity (b), the 10th-order NARMA task (c), and the delayed AND task (d). The performance for each sub-reservoir was computed as the average over 50 realizations for each one of the three tasks. In red we show two selected networks with a good size-performance compromise across the three tasks.}
\label{fig:delayed_AND}  
\end{figure}

This task is inspired by the example of associative memory of \textit{E. coli} described in Sec.~\ref{sec:intro}, in which bacteria anticipate a decrease in oxygen availability from an earlier increase in temperature \cite{tagkopoulos_predictive_2008}.
This behavior can be interpreted as a delayed AND task, in which the reservoir is stimulated by two signals separated in time, acting upon two different input nodes.
In our case, the input signal for each node consists of a series of length 600 time step, with randomly distributed pulses of unit height with a duration of 3 time steps, depicted in the top row of Fig.~\ref{fig:delayed_AND}(a). 
The ground-truth output of the task is calculated as an AND gate with inputs $I_1$ at time step $i-k$ (with $k=6$ in our case) and $I_2$ at time step $i$.
In that way, if $I_2=1$ at time step $i$ and $I_1=1$ at time step $i-k$, the system's target output is $y^{\mathrm{target}}=1$, as shown in the middle and bottom rows in Fig.~\ref{fig:delayed_AND}(a). 
The bottom row, in particular, shows the output predicted by the whole \textit{E. coli}'s reservoir (70 genes) in green. To quantify the network performance, we first binarize the output using a threshold 0.5, shown as a grey line in the bottom row of the panel.
Once binarized, the quality of the performance is evaluated by calculating the NRMSE between the binarized predicted and target output signals ($\mathrm{output}^{\mathrm{bin}}$ and $y^{\mathrm{target}}$), as been defined in Eq.~\eqref{NRMSE}.

\subsection{Network size influence on computational performance}

The response of the sub-reservoirs introduced in Section \ref{subsec:mem_cap_sub} to the three tasks defined above is shown in Figs.~\ref{fig:delayed_AND}(b-d), which considers only the networks whose performance is within 1.25 standard deviations from the mean value across all sub-reservoirs.
The plots show, as expected, that the memory capacity grows, and the error of the NARMA and delayed AND tasks decrease, as the size of the sub-graphs increases. 
We then used those results to identify two sub-reservoirs, shown as red symbols in Figs.~\ref{fig:delayed_AND}(b-d), that have a good size-performance trade-off across the three tasks considered.
These recurrent networks, whose size is $\approx 40$ nodes, have performances not much smaller than the entire reservoir, with only 57\% of its genes.
Even smaller networks can have good enough performance, although in this case only for some of the tasks.
It will be interesting to investigate in the future what topological characteristics of these small networks lead to their good performance in some of the tasks, depending on the differential characteristics of the successful task with respect to the others (e.g. the need or more or less memory, or the need to respond to nonlinear behaviors).
{Such studies could be relevant of the design of synthetic circuits with enhanced memory capacities} \cite{siuti2013synthetic,inniss2013building}.

\section{Discussion} 

Here we have studied the temporal information processing capabilities of the gene regulatory network of the bacterium \textit{Escherichia coli} within the reservoir computing framework.
{We focused on \textit{E. coli} due to the fact that the global architecture of this organism’s gene regulatory network has been systematically studied \cite{gabalda-sagarra_recurrence-based_2018}, and to the relative ease with which this bacterium can be monitored and perturbed, both genetically and chemically. 
However, the network architecture of other microorganisms and eukaryotes has similar global structural properties \cite{gabalda-sagarra_recurrence-based_2018}, and therefore we expect the results presented here to be generalizable to other living systems.}.

We concentrated on the dynamics of the recurrent core, due to its relevance in temporal information processing.
We first explored whether the dynamical regime of the reservoir affects its memory capacity. 
This was inspired by work done on artificial ESNs, where it has been seen that the performance of reservoirs are optimal at the critical point of an order-to-chaos transition \cite{legenstein_edge_2007,Morales2021}.
Our results show that the \textit{E. coli}'s reservoir has maximum memory capacity just around the onset of the transition to chaos.

We also explored whether the balance between transcriptional activation and repression is relevant in order to obtain a functional GRN under the RC framework. The repression percentage in the reservoir is 41\%. By varying this percentage artificially we saw that the optimal performance for two memory-demanding tasks (memory capacity and 10th-order NARMA) are obtained for a repression percentage near 40\%. We thus conjecture that evolution might have tuned GRNs to an optimal activation/repression ratio that maximizes memory capacity.
{It would be very interesting to validate this result, although it is certainly challenging to do this validation directly with the currently existing data. Two potential ways to do this validation, provided robust data were available, would be the following. First, one could compare different cell types in a multicellular organism, to look for substantial changes in the relative abundance of activators and repressors among cell types. In those cases, one could then correlate the activation/repression ratio with the memory capacity (or need thereof) of the corresponding cell types.
As a second potential validation method, one could look at mutant libraries designed to identify minimal genomes (in bacteria such as \textit{Mycoplasma genitalium}, for instance \cite{glass06}). Again, if the activator/repressor ratio changes systematically across these libraries, one could correlate that ratio with the memory capacity of the corresponding mutants.}

We next asked whether the local topology of GRNs contributes to their memory-encoding capabilities. To that end, we constructed a sample of $\approx$2400 sub-reservoirs from the full \textit{E. coli}'s reservoir, and aimed to find a relation between the number of motifs in the sub-reservoirs and their memory capabilities. Our results show that network motifs with recurrences (self-loops and mutual regulation) are more important for the reservoir's memory capacity than feedforward circuits, such as FFLs.
{This conclusion can be expected to hold even when other factors such as metabolic constraints are considered, as long as these factors affect gene expression globally. In that case, we expect the relative importance of the different local motifs on the resulting memory capacity to be unchanged.}

Finally, we looked for \textit{E. coli}'s sub-reservoirs with a good size-performance compromise. We identified two different sub-networks that exhibit high performance in three memory-demanding tasks: memory capacity, 10th-order NARMA task, and the more biologically plausible delayed AND task.
Taken together, our results indicate that a balanced global activation/repression ratio and local recurrences, in the form of self loops and mutual regulation circuits, grant biological networks (in particular the gene regulatory network of \textit{E. coli}) with temporal information processing capabilities.

\section*{Code availability}
All simulations shown in this article have been generated with custom-made Python code that can be obtained from \href{https://github.com/dsb-lab/memory-encoding-GRN}{https://github.com/dsb-lab/memory-encoding-GRN}.

\section*{Acknowledgments}

This work was supported by project PID2021-127311NB-I00, financed by the Spanish Ministry of Science and Innovation, the Spanish State Research Agency and the European Regional Development Fund (FEDER).
Financial support was also provided by the Maria de Maeztu Programme for Units of Excellence in R\&D (Spanish State Research Agency, project CEX2018-000792-M), and by the ICREA Academia programme.
M.S.V. was supported by a PhD fellowship from the Ag\`encia de Gesti\'o d'Ajuts Universitaris i de Recerca (AGAUR) from the Generalitat de Catalunya (grant 2021-FI-B-00408).

\bibliography{structural}


\begin{thebibliography}{47}
\ifx \bisbn   \undefined \def \bisbn  #1{ISBN #1}\fi
\ifx \binits  \undefined \def \binits#1{#1}\fi
\ifx \bauthor  \undefined \def \bauthor#1{#1}\fi
\ifx \batitle  \undefined \def \batitle#1{#1}\fi
\ifx \bjtitle  \undefined \def \bjtitle#1{#1}\fi
\ifx \bvolume  \undefined \def \bvolume#1{\textbf{#1}}\fi
\ifx \byear  \undefined \def \byear#1{#1}\fi
\ifx \bissue  \undefined \def \bissue#1{#1}\fi
\ifx \bfpage  \undefined \def \bfpage#1{#1}\fi
\ifx \blpage  \undefined \def \blpage #1{#1}\fi
\ifx \burl  \undefined \def \burl#1{\textsf{#1}}\fi
\ifx \doiurl  \undefined \def \doiurl#1{\url{https://doi.org/#1}}\fi
\ifx \betal  \undefined \def \betal{\textit{et al.}}\fi
\ifx \binstitute  \undefined \def \binstitute#1{#1}\fi
\ifx \binstitutionaled  \undefined \def \binstitutionaled#1{#1}\fi
\ifx \bctitle  \undefined \def \bctitle#1{#1}\fi
\ifx \beditor  \undefined \def \beditor#1{#1}\fi
\ifx \bpublisher  \undefined \def \bpublisher#1{#1}\fi
\ifx \bbtitle  \undefined \def \bbtitle#1{#1}\fi
\ifx \bedition  \undefined \def \bedition#1{#1}\fi
\ifx \bseriesno  \undefined \def \bseriesno#1{#1}\fi
\ifx \blocation  \undefined \def \blocation#1{#1}\fi
\ifx \bsertitle  \undefined \def \bsertitle#1{#1}\fi
\ifx \bsnm \undefined \def \bsnm#1{#1}\fi
\ifx \bsuffix \undefined \def \bsuffix#1{#1}\fi
\ifx \bparticle \undefined \def \bparticle#1{#1}\fi
\ifx \barticle \undefined \def \barticle#1{#1}\fi
\bibcommenthead
\ifx \bconfdate \undefined \def \bconfdate #1{#1}\fi
\ifx \botherref \undefined \def \botherref #1{#1}\fi
\ifx \url \undefined \def \url#1{\textsf{#1}}\fi
\ifx \bchapter \undefined \def \bchapter#1{#1}\fi
\ifx \bbook \undefined \def \bbook#1{#1}\fi
\ifx \bcomment \undefined \def \bcomment#1{#1}\fi
\ifx \oauthor \undefined \def \oauthor#1{#1}\fi
\ifx \citeauthoryear \undefined \def \citeauthoryear#1{#1}\fi
\ifx \endbibitem  \undefined \def \endbibitem {}\fi
\ifx \bconflocation  \undefined \def \bconflocation#1{#1}\fi
\ifx \arxivurl  \undefined \def \arxivurl#1{\textsf{#1}}\fi
\csname PreBibitemsHook\endcsname

\bibitem[\protect\citeauthoryear{McCulloch and
  Pitts}{1943}]{mcculloch1943logical}
\begin{barticle}
\bauthor{\bsnm{McCulloch}, \binits{W.S.}},
\bauthor{\bsnm{Pitts}, \binits{W.}}:
\batitle{A logical calculus of the ideas immanent in nervous activity}.
\bjtitle{The bulletin of mathematical biophysics}
\bvolume{5},
\bfpage{115}--\blpage{133}
(\byear{1943})
\doiurl{10.1007/BF02478259}
\end{barticle}
\endbibitem

\bibitem[\protect\citeauthoryear{Barack and Krakauer}{2021}]{barack2021two}
\begin{barticle}
\bauthor{\bsnm{Barack}, \binits{D.L.}},
\bauthor{\bsnm{Krakauer}, \binits{J.W.}}:
\batitle{Two views on the cognitive brain}.
\bjtitle{Nature Reviews Neuroscience}
\bvolume{22}(\bissue{6}),
\bfpage{359}--\blpage{371}
(\byear{2021})
\doiurl{10.1038/s41583-021-00448-6}
\end{barticle}
\endbibitem

\bibitem[\protect\citeauthoryear{Gunawardena}{2022}]{Gunawardena2022}
\begin{barticle}
\bauthor{\bsnm{Gunawardena}, \binits{J.}}:
\batitle{Learning outside the brain: Integrating cognitive science and systems
  biology}.
\bjtitle{Proceedings of the IEEE}
\bvolume{110}(\bissue{5}),
\bfpage{590}--\blpage{612}
(\byear{2022})
\doiurl{10.1109/JPROC.2022.3162791}
\end{barticle}
\endbibitem

\bibitem[\protect\citeauthoryear{Lee et~al.}{2002}]{lee_transcriptional_2002}
\begin{barticle}
\bauthor{\bsnm{Lee}, \binits{T.I.}},
\bauthor{\bsnm{Rinaldi}, \binits{N.J.}},
\bauthor{\bsnm{Robert}, \binits{F.}},
\bauthor{\bsnm{Odom}, \binits{D.T.}},
\bauthor{\bsnm{Bar-Joseph}, \binits{Z.}},
\bauthor{\bsnm{Gerber}, \binits{G.K.}},
\bauthor{\bsnm{Hannett}, \binits{N.M.}},
\bauthor{\bsnm{Harbison}, \binits{C.T.}},
\bauthor{\bsnm{Thompson}, \binits{C.M.}},
\bauthor{\bsnm{Simon}, \binits{I.}},
\bauthor{\bsnm{Zeitlinger}, \binits{J.}},
\bauthor{\bsnm{Jennings}, \binits{E.G.}},
\bauthor{\bsnm{Murray}, \binits{H.L.}},
\bauthor{\bsnm{Gordon}, \binits{D.B.}},
\bauthor{\bsnm{Ren}, \binits{B.}},
\bauthor{\bsnm{Wyrick}, \binits{J.J.}},
\bauthor{\bsnm{Tagne}, \binits{J.-B.}},
\bauthor{\bsnm{Volkert}, \binits{T.L.}},
\bauthor{\bsnm{Fraenkel}, \binits{E.}},
\bauthor{\bsnm{Gifford}, \binits{D.K.}},
\bauthor{\bsnm{Young}, \binits{R.A.}}:
\batitle{Transcriptional {Regulatory} {Networks} in {Saccharomyces}
  cerevisiae}.
\bjtitle{Science}
\bvolume{298}(\bissue{5594}),
\bfpage{799}--\blpage{804}
(\byear{2002})
\doiurl{10.1126/science.1075090}
\end{barticle}
\endbibitem

\bibitem[\protect\citeauthoryear{Martinez-Antonio and
  Collado-Vides}{2003}]{martinez-antonio_identifying_2003}
\begin{barticle}
\bauthor{\bsnm{Martinez-Antonio}, \binits{A.}},
\bauthor{\bsnm{Collado-Vides}, \binits{J.}}:
\batitle{Identifying global regulators in transcriptional regulatory networks
  in bacteria}.
\bjtitle{Current Opinion in Microbiology}
\bvolume{6}(\bissue{5}),
\bfpage{482}--\blpage{489}
(\byear{2003})
\doiurl{10.1016/j.mib.2003.09.002}
\end{barticle}
\endbibitem

\bibitem[\protect\citeauthoryear{Mjolsness
  et~al.}{1991}]{mjolsness1991connectionist}
\begin{barticle}
\bauthor{\bsnm{Mjolsness}, \binits{E.}},
\bauthor{\bsnm{Sharp}, \binits{D.H.}},
\bauthor{\bsnm{Reinitz}, \binits{J.}}:
\batitle{A connectionist model of development}.
\bjtitle{Journal of Theoretical Biology}
\bvolume{152}(\bissue{4}),
\bfpage{429}--\blpage{453}
(\byear{1991})
\doiurl{10.1016/S0022-5193(05)80391-1}
\end{barticle}
\endbibitem

\bibitem[\protect\citeauthoryear{Tagkopoulos
  et~al.}{2008}]{tagkopoulos_predictive_2008}
\begin{barticle}
\bauthor{\bsnm{Tagkopoulos}, \binits{I.}},
\bauthor{\bsnm{Liu}, \binits{Y.-C.}},
\bauthor{\bsnm{Tavazoie}, \binits{S.}}:
\batitle{Predictive {Behavior} {Within} {Microbial} {Genetic} {Networks}}.
\bjtitle{Science}
\bvolume{320}(\bissue{5881}),
\bfpage{1313}--\blpage{1317}
(\byear{2008})
\doiurl{10.1126/science.1154456}
\end{barticle}
\endbibitem

\bibitem[\protect\citeauthoryear{Schild et~al.}{2007}]{schild_genes_2007}
\begin{barticle}
\bauthor{\bsnm{Schild}, \binits{S.}},
\bauthor{\bsnm{Tamayo}, \binits{R.}},
\bauthor{\bsnm{Nelson}, \binits{E.J.}},
\bauthor{\bsnm{Qadri}, \binits{F.}},
\bauthor{\bsnm{Calderwood}, \binits{S.B.}},
\bauthor{\bsnm{Camilli}, \binits{A.}}:
\batitle{Genes {Induced} {Late} in {Infection} {Increase} {Fitness} of {Vibrio}
  cholerae after {Release} into the {Environment}}.
\bjtitle{Cell Host \& Microbe}
\bvolume{2}(\bissue{4}),
\bfpage{264}--\blpage{277}
(\byear{2007})
\doiurl{10.1016/j.chom.2007.09.004}
\end{barticle}
\endbibitem

\bibitem[\protect\citeauthoryear{Wolf et~al.}{2008}]{wolf_memory_2008}
\begin{barticle}
\bauthor{\bsnm{Wolf}, \binits{D.M.}},
\bauthor{\bsnm{Fontaine-Bodin}, \binits{L.}},
\bauthor{\bsnm{Bischofs}, \binits{I.}},
\bauthor{\bsnm{Price}, \binits{G.}},
\bauthor{\bsnm{Keasling}, \binits{J.}},
\bauthor{\bsnm{Arkin}, \binits{A.P.}}:
\batitle{Memory in {Microbes}: {Quantifying} {History}-{Dependent} {Behavior}
  in a {Bacterium}}.
\bjtitle{PLOS ONE}
\bvolume{3}(\bissue{2}),
\bfpage{1700}
(\byear{2008})
\doiurl{10.1371/journal.pone.0001700}
\end{barticle}
\endbibitem

\bibitem[\protect\citeauthoryear{Brown and Milner}{2003}]{brown_legacy_2003}
\begin{barticle}
\bauthor{\bsnm{Brown}, \binits{R.E.}},
\bauthor{\bsnm{Milner}, \binits{P.M.}}:
\batitle{The legacy of {Donald} {O}. {Hebb}: more than the {Hebb} {Synapse}}.
\bjtitle{Nature Reviews Neuroscience}
\bvolume{4}(\bissue{12}),
\bfpage{1013}--\blpage{1019}
(\byear{2003})
\doiurl{10.1038/nrn1257}
\end{barticle}
\endbibitem

\bibitem[\protect\citeauthoryear{Casal et~al.}{2020}]{Casal2020}
\begin{botherref}
\oauthor{\bsnm{Casal}, \binits{M.A.}},
\oauthor{\bsnm{Galella}, \binits{S.}},
\oauthor{\bsnm{Vilarroya}, \binits{O.}},
\oauthor{\bsnm{Garcia-Ojalvo}, \binits{J.}}:
Soft-wired long-term memory in a natural recurrent neuronal network.
Chaos: An Interdisciplinary Journal of Nonlinear Science
\textbf{30}(6)
(2020)
\doiurl{10.1063/5.0009709}
\end{botherref}
\endbibitem

\bibitem[\protect\citeauthoryear{Sussillo and
  Abbott}{2009}]{sussillo2009generating}
\begin{barticle}
\bauthor{\bsnm{Sussillo}, \binits{D.}},
\bauthor{\bsnm{Abbott}, \binits{L.F.}}:
\batitle{Generating coherent patterns of activity from chaotic neural
  networks}.
\bjtitle{Neuron}
\bvolume{63}(\bissue{4}),
\bfpage{544}--\blpage{557}
(\byear{2009})
\doiurl{10.1016/j.neuron.2009.07.018}
\end{barticle}
\endbibitem

\bibitem[\protect\citeauthoryear{Buonomano and
  Maass}{2009}]{buonomano_state-dependent_2009}
\begin{barticle}
\bauthor{\bsnm{Buonomano}, \binits{D.V.}},
\bauthor{\bsnm{Maass}, \binits{W.}}:
\batitle{State-dependent computations: spatiotemporal processing in cortical
  networks}.
\bjtitle{Nature Reviews Neuroscience}
\bvolume{10}(\bissue{2}),
\bfpage{113}--\blpage{125}
(\byear{2009})
\doiurl{10.1038/nrn2558}
\end{barticle}
\endbibitem

\bibitem[\protect\citeauthoryear{Verstraeten
  et~al.}{2007}]{verstraeten_experimental_2007}
\begin{barticle}
\bauthor{\bsnm{Verstraeten}, \binits{D.}},
\bauthor{\bsnm{Schrauwen}, \binits{B.}},
\bauthor{\bsnm{D’Haene}, \binits{M.}},
\bauthor{\bsnm{Stroobandt}, \binits{D.}}:
\batitle{An experimental unification of reservoir computing methods}.
\bjtitle{Neural Networks}
\bvolume{20}(\bissue{3}),
\bfpage{391}--\blpage{403}
(\byear{2007})
\doiurl{10.1016/j.neunet.2007.04.003}
\end{barticle}
\endbibitem

\bibitem[\protect\citeauthoryear{Jaeger}{2001}]{jaeger_echo_2001}
\begin{barticle}
\bauthor{\bsnm{Jaeger}, \binits{H.}}:
\batitle{The “echo state” approach to analysing and training recurrent
  neural networks}.
\bjtitle{GMD Technical Report}
\bvolume{148},
\bfpage{1}
(\byear{2001})
\doiurl{10.24406/publica-fhg-291111}
\end{barticle}
\endbibitem

\bibitem[\protect\citeauthoryear{Maass et~al.}{2002}]{maass_real-time_2002}
\begin{barticle}
\bauthor{\bsnm{Maass}, \binits{W.}},
\bauthor{\bsnm{Natschläger}, \binits{T.}},
\bauthor{\bsnm{Markram}, \binits{H.}}:
\batitle{Real-{Time} {Computing} {Without} {Stable} {States}: {A} {New}
  {Framework} for {Neural} {Computation} {Based} on {Perturbations}}.
\bjtitle{Neural Computation}
\bvolume{14}(\bissue{11}),
\bfpage{2531}--\blpage{2560}
(\byear{2002})
\doiurl{10.1162/089976602760407955}
\end{barticle}
\endbibitem

\bibitem[\protect\citeauthoryear{Jones et~al.}{2007}]{jones_is_2007}
\begin{bchapter}
\bauthor{\bsnm{Jones}, \binits{B.}},
\bauthor{\bsnm{Stekel}, \binits{D.}},
\bauthor{\bsnm{Rowe}, \binits{J.}},
\bauthor{\bsnm{Fernando}, \binits{C.}}:
\bctitle{Is there a {Liquid} {State} {Machine} in the {Bacterium} {Escherichia}
  {Coli}?}
In: \bbtitle{2007 {IEEE} {Symposium} on {Artificial} {Life}},
pp. \bfpage{187}--\blpage{191}
(\byear{2007}).
\doiurl{10.1109/ALIFE.2007.367795}
\end{bchapter}
\endbibitem

\bibitem[\protect\citeauthoryear{Gabalda-Sagarra
  et~al.}{2018}]{gabalda-sagarra_recurrence-based_2018}
\begin{barticle}
\bauthor{\bsnm{Gabalda-Sagarra}, \binits{M.}},
\bauthor{\bsnm{Carey}, \binits{L.B.}},
\bauthor{\bsnm{Garcia-Ojalvo}, \binits{J.}}:
\batitle{Recurrence-based information processing in gene regulatory networks}.
\bjtitle{Chaos: An Interdisciplinary Journal of Nonlinear Science}
\bvolume{28}(\bissue{10}),
\bfpage{106313}
(\byear{2018})
\doiurl{10.1063/1.5039861}
\end{barticle}
\endbibitem

\bibitem[\protect\citeauthoryear{Van~Vreeswijk and
  Sompolinsky}{1996}]{van1996chaos}
\begin{barticle}
\bauthor{\bsnm{Van~Vreeswijk}, \binits{C.}},
\bauthor{\bsnm{Sompolinsky}, \binits{H.}}:
\batitle{Chaos in neuronal networks with balanced excitatory and inhibitory
  activity}.
\bjtitle{Science}
\bvolume{274}(\bissue{5293}),
\bfpage{1724}--\blpage{1726}
(\byear{1996})
\doiurl{10.1126/science.274.5293.172}
\end{barticle}
\endbibitem

\bibitem[\protect\citeauthoryear{Keseler et~al.}{2011}]{keseler_ecocyc_2011}
\begin{barticle}
\bauthor{\bsnm{Keseler}, \binits{I.M.}},
\bauthor{\bsnm{Collado-Vides}, \binits{J.}},
\bauthor{\bsnm{Santos-Zavaleta}, \binits{A.}},
\bauthor{\bsnm{Peralta-Gil}, \binits{M.}},
\bauthor{\bsnm{Gama-Castro}, \binits{S.}},
\bauthor{\bsnm{Muñiz-Rascado}, \binits{L.}},
\bauthor{\bsnm{Bonavides-Martinez}, \binits{C.}},
\bauthor{\bsnm{Paley}, \binits{S.}},
\bauthor{\bsnm{Krummenacker}, \binits{M.}},
\bauthor{\bsnm{Altman}, \binits{T.}},
\bauthor{\bsnm{Kaipa}, \binits{P.}},
\bauthor{\bsnm{Spaulding}, \binits{A.}},
\bauthor{\bsnm{Pacheco}, \binits{J.}},
\bauthor{\bsnm{Latendresse}, \binits{M.}},
\bauthor{\bsnm{Fulcher}, \binits{C.}},
\bauthor{\bsnm{Sarker}, \binits{M.}},
\bauthor{\bsnm{Shearer}, \binits{A.G.}},
\bauthor{\bsnm{Mackie}, \binits{A.}},
\bauthor{\bsnm{Paulsen}, \binits{I.}},
\bauthor{\bsnm{Gunsalus}, \binits{R.P.}},
\bauthor{\bsnm{Karp}, \binits{P.D.}}:
\batitle{{EcoCyc}: a comprehensive database of {Escherichia} coli biology}.
\bjtitle{Nucleic Acids Research}
\bvolume{39}(\bissue{suppl\_1}),
\bfpage{583}--\blpage{590}
(\byear{2011})
\doiurl{10.1093/nar/gkq1143}
\end{barticle}
\endbibitem

\bibitem[\protect\citeauthoryear{Lukoševičius and
  Jaeger}{2009}]{lukosevicius_reservoir_2009}
\begin{barticle}
\bauthor{\bsnm{Lukoševičius}, \binits{M.}},
\bauthor{\bsnm{Jaeger}, \binits{H.}}:
\batitle{Reservoir computing approaches to recurrent neural network training}.
\bjtitle{Computer Science Review}
\bvolume{3}(\bissue{3}),
\bfpage{127}--\blpage{149}
(\byear{2009})
\doiurl{10.1016/j.cosrev.2009.03.005}
\end{barticle}
\endbibitem

\bibitem[\protect\citeauthoryear{Wyffels et~al.}{2008}]{wyffels_stable_2008}
\begin{bchapter}
\bauthor{\bsnm{Wyffels}, \binits{F.}},
\bauthor{\bsnm{Schrauwen}, \binits{B.}},
\bauthor{\bsnm{Stroobandt}, \binits{D.}}:
\bctitle{Stable {Output} {Feedback} in {Reservoir} {Computing} {Using} {Ridge}
  {Regression}}.
In: \beditor{\bsnm{Kůrková}, \binits{V.}},
\beditor{\bsnm{Neruda}, \binits{R.}},
\beditor{\bsnm{Koutník}, \binits{J.}} (eds.)
\bbtitle{Artificial {Neural} {Networks} - {ICANN} 2008}.
\bsertitle{Lecture {Notes} in {Computer} {Science}},
pp. \bfpage{808}--\blpage{817}.
\bpublisher{Springer},
\blocation{Berlin, Heidelberg}
(\byear{2008}).
\doiurl{10.1007/978-3-540-87536-9}
\end{bchapter}
\endbibitem

\bibitem[\protect\citeauthoryear{Watson and
  Szathm{\'a}ry}{2016}]{watson2016can}
\begin{barticle}
\bauthor{\bsnm{Watson}, \binits{R.A.}},
\bauthor{\bsnm{Szathm{\'a}ry}, \binits{E.}}:
\batitle{How can evolution learn?}
\bjtitle{Trends in ecology \& evolution}
\bvolume{31}(\bissue{2}),
\bfpage{147}--\blpage{157}
(\byear{2016})
\doiurl{10.1016/j.tree.2015.11.009}
\end{barticle}
\endbibitem

\bibitem[\protect\citeauthoryear{Bertschinger and
  Natschläger}{2004}]{bertschinger_real-time_2004}
\begin{barticle}
\bauthor{\bsnm{Bertschinger}, \binits{N.}},
\bauthor{\bsnm{Natschläger}, \binits{T.}}:
\batitle{Real-{Time} {Computation} at the {Edge} of {Chaos} in {Recurrent}
  {Neural} {Networks}}.
\bjtitle{Neural Computation}
\bvolume{16}(\bissue{7}),
\bfpage{1413}--\blpage{1436}
(\byear{2004})
\doiurl{10.1162/089976604323057443}
\end{barticle}
\endbibitem

\bibitem[\protect\citeauthoryear{Legenstein and
  Maass}{2007}]{legenstein_edge_2007}
\begin{barticle}
\bauthor{\bsnm{Legenstein}, \binits{R.}},
\bauthor{\bsnm{Maass}, \binits{W.}}:
\batitle{Edge of chaos and prediction of computational performance for neural
  circuit models}.
\bjtitle{Neural Networks}
\bvolume{20}(\bissue{3}),
\bfpage{323}--\blpage{334}
(\byear{2007})
\doiurl{10.1016/j.neunet.2007.04.017}
\end{barticle}
\endbibitem

\bibitem[\protect\citeauthoryear{Morales and Muñoz}{2021}]{Morales2021}
\begin{barticle}
\bauthor{\bsnm{Morales}, \binits{G.B.}},
\bauthor{\bsnm{Muñoz}, \binits{M.A.}}:
\batitle{Optimal input representation in neural systems at the edge of chaos}.
\bjtitle{Biology}
\bvolume{10}(\bissue{8}),
\bfpage{702}
(\byear{2021})
\doiurl{10.3390/biology10080702}
\end{barticle}
\endbibitem

\bibitem[\protect\citeauthoryear{Vidal-Saez et~al.}{2024}]{vidal_BBRC_24}
\begin{barticle}
\bauthor{\bsnm{Vidal-Saez}, \binits{M.S.}},
\bauthor{\bsnm{Vilarroya}, \binits{O.}},
\bauthor{\bsnm{Garcia-Ojalvo}, \binits{J.}}:
\batitle{Biological computation through recurrence}.
\bjtitle{Biochemical and Biophysical Research Communications}
\bvolume{728},
\bfpage{150301}
(\byear{2024})
\doiurl{10.1016/j.bbrc.2024.150301}
\end{barticle}
\endbibitem

\bibitem[\protect\citeauthoryear{Beggs}{2007}]{beggs_criticality_2007}
\begin{barticle}
\bauthor{\bsnm{Beggs}, \binits{J.M.}}:
\batitle{The criticality hypothesis: how local cortical networks might optimize
  information processing}.
\bjtitle{Philosophical Transactions of the Royal Society A: Mathematical,
  Physical and Engineering Sciences}
\bvolume{366}(\bissue{1864}),
\bfpage{329}--\blpage{343}
(\byear{2007})
\doiurl{10.1098/rsta.2007.2092}
\end{barticle}
\endbibitem

\bibitem[\protect\citeauthoryear{Beggs and Plenz}{2003}]{beggs_neuronal_2003}
\begin{barticle}
\bauthor{\bsnm{Beggs}, \binits{J.M.}},
\bauthor{\bsnm{Plenz}, \binits{D.}}:
\batitle{Neuronal {Avalanches} in {Neocortical} {Circuits}}.
\bjtitle{Journal of Neuroscience}
\bvolume{23}(\bissue{35}),
\bfpage{11167}--\blpage{11177}
(\byear{2003})
\doiurl{10.1523/JNEUROSCI.23-35-11167.2003}
\end{barticle}
\endbibitem

\bibitem[\protect\citeauthoryear{Chialvo}{2004}]{r_chialvo_critical_2004}
\begin{barticle}
\bauthor{\bsnm{Chialvo}, \binits{D.R.}}:
\batitle{Critical brain networks}.
\bjtitle{Physica A: Statistical Mechanics and its Applications}
\bvolume{340}(\bissue{4}),
\bfpage{756}--\blpage{765}
(\byear{2004})
\doiurl{10.1016/j.physa.2004.05.064}
\end{barticle}
\endbibitem

\bibitem[\protect\citeauthoryear{Vidiella et~al.}{2021}]{criticality_blai}
\begin{barticle}
\bauthor{\bsnm{Vidiella}, \binits{B.}},
\bauthor{\bsnm{Guillamon}, \binits{A.}},
\bauthor{\bsnm{Sardany{\'e}s}, \binits{J.}},
\bauthor{\bsnm{Maull}, \binits{V.}},
\bauthor{\bsnm{Pla}, \binits{J.}},
\bauthor{\bsnm{Conde}, \binits{N.}},
\bauthor{\bsnm{Sol{\'e}}, \binits{R.}}:
\batitle{Engineering self-organized criticality in living cells}.
\bjtitle{Nature Communications}
\bvolume{12}(\bissue{1}),
\bfpage{4415}
(\byear{2021})
\doiurl{10.1038/s41467-021-24695-4}
\end{barticle}
\endbibitem

\bibitem[\protect\citeauthoryear{Jaeger}{2001}]{jaeger_short_2001}
\begin{barticle}
\bauthor{\bsnm{Jaeger}, \binits{H.}}:
\batitle{Short term memory in echo state networks}.
\bjtitle{GMD Technical Report}
\bvolume{152},
\bfpage{1}
(\byear{2001})
\doiurl{10.24406/publica-fhg-291107}
\end{barticle}
\endbibitem

\bibitem[\protect\citeauthoryear{Yildiz et~al.}{2012}]{yildiz_re-visiting_2012}
\begin{barticle}
\bauthor{\bsnm{Yildiz}, \binits{I.B.}},
\bauthor{\bsnm{Jaeger}, \binits{H.}},
\bauthor{\bsnm{Kiebel}, \binits{S.J.}}:
\batitle{Re-visiting the echo state property}.
\bjtitle{Neural Networks}
\bvolume{35},
\bfpage{1}--\blpage{9}
(\byear{2012})
\doiurl{10.1016/j.neunet.2012.07.005}
\end{barticle}
\endbibitem

\bibitem[\protect\citeauthoryear{Boedecker
  et~al.}{2012}]{boedecker_information_2012}
\begin{barticle}
\bauthor{\bsnm{Boedecker}, \binits{J.}},
\bauthor{\bsnm{Obst}, \binits{O.}},
\bauthor{\bsnm{Lizier}, \binits{J.T.}},
\bauthor{\bsnm{Mayer}, \binits{N.M.}},
\bauthor{\bsnm{Asada}, \binits{M.}}:
\batitle{Information processing in echo state networks at the edge of chaos}.
\bjtitle{Theory in Biosciences}
\bvolume{131}(\bissue{3}),
\bfpage{205}--\blpage{213}
(\byear{2012})
\doiurl{10.1007/s12064-011-0146-8}
\end{barticle}
\endbibitem

\bibitem[\protect\citeauthoryear{Derrida and
  Pomeau}{1986}]{derrida_random_1986}
\begin{barticle}
\bauthor{\bsnm{Derrida}, \binits{B.}},
\bauthor{\bsnm{Pomeau}, \binits{Y.}}:
\batitle{Random {Networks} of {Automata}: {A} {Simple} {Annealed}
  {Approximation}}.
\bjtitle{Europhysics Letters}
\bvolume{1}(\bissue{2}),
\bfpage{45}
(\byear{1986})
\doiurl{10.1209/0295-5075/1/2/001}
\end{barticle}
\endbibitem

\bibitem[\protect\citeauthoryear{Sprott}{2003}]{Sprott2003-ln}
\begin{bbook}
\bauthor{\bsnm{Sprott}, \binits{J.C.}}:
\bbtitle{Chaos and Time-series Analysis}.
\bpublisher{Oxford University Press},
\blocation{London, England}
(\byear{2003}).
\doiurl{10.5860/choice.41-3492}
\end{bbook}
\endbibitem

\bibitem[\protect\citeauthoryear{Deco et~al.}{2014}]{deco_how_2014}
\begin{barticle}
\bauthor{\bsnm{Deco}, \binits{G.}},
\bauthor{\bsnm{Ponce-Alvarez}, \binits{A.}},
\bauthor{\bsnm{Hagmann}, \binits{P.}},
\bauthor{\bsnm{Romani}, \binits{G.L.}},
\bauthor{\bsnm{Mantini}, \binits{D.}},
\bauthor{\bsnm{Corbetta}, \binits{M.}}:
\batitle{How {Local} {Excitation}–{Inhibition} {Ratio} {Impacts} the {Whole}
  {Brain} {Dynamics}}.
\bjtitle{Journal of Neuroscience}
\bvolume{34}(\bissue{23}),
\bfpage{7886}--\blpage{7898}
(\byear{2014})
\doiurl{10.1523/JNEUROSCI.5068-13.2014}
\end{barticle}
\endbibitem

\bibitem[\protect\citeauthoryear{Ghatak et~al.}{2021}]{ghatak_novel_2021}
\begin{barticle}
\bauthor{\bsnm{Ghatak}, \binits{S.}},
\bauthor{\bsnm{Talantova}, \binits{M.}},
\bauthor{\bsnm{McKercher}, \binits{S.R.}},
\bauthor{\bsnm{Lipton}, \binits{S.A.}}:
\batitle{Novel {Therapeutic} {Approach} for {Excitatory}/{Inhibitory}
  {Imbalance} in {Neurodevelopmental} and {Neurodegenerative} {Diseases}}.
\bjtitle{Annual Review of Pharmacology and Toxicology}
\bvolume{61}(\bissue{1}),
\bfpage{701}--\blpage{721}
(\byear{2021})
\doiurl{10.1146/annurev-pharmtox-032320-015420}
\end{barticle}
\endbibitem

\bibitem[\protect\citeauthoryear{Kirischuk}{2022}]{kirischuk_keeping_2022}
\begin{barticle}
\bauthor{\bsnm{Kirischuk}, \binits{S.}}:
\batitle{Keeping {Excitation}–{Inhibition} {Ratio} in {Balance}}.
\bjtitle{International Journal of Molecular Sciences}
\bvolume{23}(\bissue{10}),
\bfpage{5746}
(\byear{2022})
\doiurl{10.3390/ijms23105746}
\end{barticle}
\endbibitem

\bibitem[\protect\citeauthoryear{Jaeger}{2002}]{jaeger_adaptive_2002}
\begin{botherref}
\oauthor{\bsnm{Jaeger}, \binits{H.}}:
Adaptive nonlinear system identification with echo state networks.
Advances in neural information processing systems
\textbf{15}
(2002)
\end{botherref}
\endbibitem

\bibitem[\protect\citeauthoryear{Atiya and Parlos}{2000}]{atiya_new_2000}
\begin{barticle}
\bauthor{\bsnm{Atiya}, \binits{A.F.}},
\bauthor{\bsnm{Parlos}, \binits{A.G.}}:
\batitle{New results on recurrent network training: unifying the algorithms and
  accelerating convergence}.
\bjtitle{IEEE Transactions on Neural Networks}
\bvolume{11}(\bissue{3}),
\bfpage{697}--\blpage{709}
(\byear{2000})
\doiurl{10.1109/72.846741}
\end{barticle}
\endbibitem

\bibitem[\protect\citeauthoryear{Alon}{2019}]{alon_introduction_2019}
\begin{bbook}
\bauthor{\bsnm{Alon}, \binits{U.}}:
\bbtitle{An {Introduction} to {Systems} {Biology}}.
\bpublisher{CRC Press},
\blocation{New York}
(\byear{2019}).
\doiurl{10.1201/9781420011432}
\end{bbook}
\endbibitem

\bibitem[\protect\citeauthoryear{Milo et~al.}{2002}]{milo_network_2002}
\begin{barticle}
\bauthor{\bsnm{Milo}, \binits{R.}},
\bauthor{\bsnm{Shen-Orr}, \binits{S.}},
\bauthor{\bsnm{Itzkovitz}, \binits{S.}},
\bauthor{\bsnm{Kashtan}, \binits{N.}},
\bauthor{\bsnm{Chklovskii}, \binits{D.}},
\bauthor{\bsnm{Alon}, \binits{U.}}:
\batitle{Network {Motifs}: {Simple} {Building} {Blocks} of {Complex}
  {Networks}}.
\bjtitle{Science}
\bvolume{298}(\bissue{5594}),
\bfpage{824}--\blpage{827}
(\byear{2002})
\doiurl{10.1126/science.298.5594.824}
\end{barticle}
\endbibitem

\bibitem[\protect\citeauthoryear{Shen-Orr et~al.}{2002}]{shen-orr_network_2002}
\begin{barticle}
\bauthor{\bsnm{Shen-Orr}, \binits{S.S.}},
\bauthor{\bsnm{Milo}, \binits{R.}},
\bauthor{\bsnm{Mangan}, \binits{S.}},
\bauthor{\bsnm{Alon}, \binits{U.}}:
\batitle{Network motifs in the transcriptional regulation network of
  {Escherichia} coli}.
\bjtitle{Nature Genetics}
\bvolume{31}(\bissue{1}),
\bfpage{64}--\blpage{68}
(\byear{2002})
\doiurl{10.1038/ng881}
\end{barticle}
\endbibitem

\bibitem[\protect\citeauthoryear{Siuti et~al.}{2013}]{siuti2013synthetic}
\begin{barticle}
\bauthor{\bsnm{Siuti}, \binits{P.}},
\bauthor{\bsnm{Yazbek}, \binits{J.}},
\bauthor{\bsnm{Lu}, \binits{T.K.}}:
\batitle{Synthetic circuits integrating logic and memory in living cells}.
\bjtitle{Nature biotechnology}
\bvolume{31}(\bissue{5}),
\bfpage{448}--\blpage{452}
(\byear{2013})
\doiurl{10.1038/nbt.2510}
\end{barticle}
\endbibitem

\bibitem[\protect\citeauthoryear{Inniss and Silver}{2013}]{inniss2013building}
\begin{barticle}
\bauthor{\bsnm{Inniss}, \binits{M.C.}},
\bauthor{\bsnm{Silver}, \binits{P.A.}}:
\batitle{Building synthetic memory}.
\bjtitle{Current Biology}
\bvolume{23}(\bissue{17}),
\bfpage{812}--\blpage{816}
(\byear{2013})
\doiurl{10.1016/j.cub.2013.06.047}
\end{barticle}
\endbibitem

\bibitem[\protect\citeauthoryear{Glass et~al.}{2006}]{glass06}
\begin{barticle}
\bauthor{\bsnm{Glass}, \binits{J.I.}},
\bauthor{\bsnm{Assad-Garcia}, \binits{N.}},
\bauthor{\bsnm{Alperovich}, \binits{N.}},
\bauthor{\bsnm{Yooseph}, \binits{S.}},
\bauthor{\bsnm{Lewis}, \binits{M.R.}},
\bauthor{\bsnm{Maruf}, \binits{M.}},
\bauthor{\bsnm{Hutchison}, \binits{C.A.}},
\bauthor{\bsnm{Smith}, \binits{H.O.}},
\bauthor{\bsnm{Venter}, \binits{J.C.}}:
\batitle{Essential genes of a minimal bacterium}.
\bjtitle{Proceedings of the National Academy of Sciences}
\bvolume{103}(\bissue{2}),
\bfpage{425}--\blpage{430}
(\byear{2006})
\doiurl{10.1073/pnas.0510013103}
\end{barticle}
\endbibitem

\end{thebibliography}

\end{document}